\documentclass{ws-procs9x6}

\begin{document}

\title{PRECISE DETERMINATIONS OF THE CHARM QUARK MASS}

\author{Matthias Steinhauser}

\address{Institut f{\"u}r Theoretische Teilchenphysik,
  Universit{\"a}t Karlsruhe (TH),\\ 76128 Karlsruhe, Germany\\
  E-mail: matthias.steinhauser@uka.de}

\begin{abstract}
In this contribution two recent analyses for the extraction of the
charm quark mass are discussed. Although they rely on completely 
different experimental and theoretical input the two methods provide
the same final results for the charm quark mass 
and have an uncertainty of about 1\%.
\end{abstract}


\bodymatter


\section{Introduction}

There has been an enormous progress in the determination of the quark
masses in the recent years due to improved experimental results, many
high-order calculations in perturbative QCD and precise lattice
simulations~\cite{Yao:2006px}.
In this contribution we describe two recent analyses
which lead to the most precise results for the $\overline{\rm MS}$ charm quark
mass. 

The first method~\cite{Novikov:1977dq,Kuhn:2001dm,Kuhn:2007vp}
 is based on four-loop
perturbative calculations for the moments of the vector correlator
which are combined with moments extracted from
precise experimental input for the total hadronic cross section in 
electron positron collisions.

Also the second method~\cite{Allison:2008xk}
relies on four-loop calculations, however, for
the pseudo-scalar rather than for the vector current correlator.
It is combined with data obtained from simulations on the lattice with
dynamical charm quarks. The latter are tuned such that the mass splitting
between the $\Upsilon^\prime$ and $\Upsilon$ and the meson masses 
$m_\pi^2$, $2 m_K^2-m_\pi^2$, $m_{\eta_c}$ and $m_\Upsilon$ are correctly
reproduced. Thus the underlying experimental data are completely different from
the first approach.


\section{\label{sec::ve}$R(s)$ and perturbative QCD}

The basic object for the first method is the
total hadronic cross section in $e^+e^-$ annihilation. Normalized to the 
production cross section of a muon pair it defines the quantity
\begin{eqnarray}
  R(s) &=& \frac{\sigma(e^+e^-\to\mbox{hadrons})}{\sigma_{\rm pt}}
  \,,
\end{eqnarray}
where $\sigma_{\rm pt} = 4\pi\alpha^2/(3s)$.

\begin{figure}[t]
  \begin{center}
    \begin{tabular}{c}
      \leavevmode
      \epsfxsize=\textwidth
      \epsffile{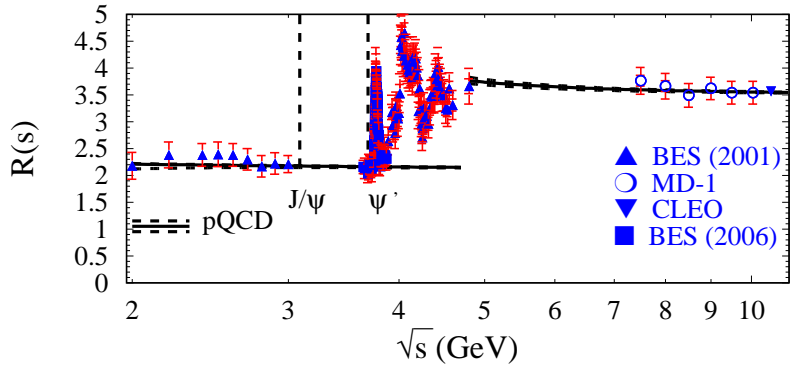}
    \end{tabular}
  \end{center}
  \caption{\label{fig::R}$R(s)$ around the charm threshold region. 
    The solid line corresponds to the theoretical prediction. The
    uncertainties, which are indicated by the dashed curves, are obtained
    from the variation of the input parameters and of $\mu$.
    The inner and outer error bars give the statistical
    and systematical uncertainty, respectively.
    Next to the data from BES~\cite{Bai:2001ct,Ablikim:2006mb}
    we also show the results form MD-1~\cite{Blinov:1993fw} and
    CLEO~\cite{Ammar:1997sk}. The narrow resonances are indicated by dashed
    lines. 
          }
\end{figure}

A compilation of the experimental data contributing to $R(s)$ in the
charm region can be found in Fig.~\ref{fig::R}. For our analysis it
is of particular importance to have precise values for the electronic
widths of the narrow resonances $J/\Psi$ and $\Psi^\prime$ which have
been measured by various experiments~\cite{Yao:2006px}. Furthermore,
we rely on the excellent data provided by the BES
collaboration~\cite{Bai:2001ct,Ablikim:2006mb} in the
region between 3.73~GeV (which is the onset of $D$ meson
production) and about 5~GeV which marks the end point of the strong
variations of $R(s)$. Above 5~GeV $R(s)$ is basically flat and can be described
very well within perturbative QCD taking into account charm quark effects. Thus
in this region we use {\tt rhad}~\cite{Harlander:2002ur}, a {\tt fortran}
program containing all state-of-the-art radiative corrections to $R(s)$
since between 5~GeV and 7~GeV no reliable data is available.

Since we are interested in the extraction of the charm quark mass we
have to consider the part of $R(s)$ which corresponds to 
the production of charm quarks, usually denoted by
$R_c(s)$. $R_c(s)$ is used to compute the so-called experimental
moments through
\begin{eqnarray}
  {\cal M}^{\rm exp}_n &\equiv& \int \frac{{\rm d}s}{s^{n+1}} R_c(s)
  \,.
  \label{eq::Mexp}
\end{eqnarray}
It is clear that in order to perform the integration in Eq.~(\ref{eq::Mexp})
one has to subtract the contributions from the three light quarks. 
This has to be done in a careful manner which is described in detail
in Ref.~\cite{Kuhn:2007vp}.

The theoretical counterpart to Eq.~(\ref{eq::Mexp}) is given by
\begin{eqnarray}
  {\cal M}_n^{\rm th} &=& \left(\frac{1}{4 m_c^2}\right)^n \bar{C}_n
  \,.
  \label{eq::Mth}
\end{eqnarray}
where the $\bar{C}_n$ are obtained from the Taylor coefficients of the 
photon polarization function for small external momentum.

Low moments are perturbative
and have long been known through three-loop order
~\cite{Chetyrkin:1995ii,Chetyrkin:1996cf,Chetyrkin:1997mb}
(see Ref.~\cite{Boughezal:2006uu,Maier:2007yn} for moments up to $n=30$).
More recently also the four-loop contribution for
$n=1$~\cite{Chetyrkin:2006xg,Boughezal:2006px} and $n=2$ could be
evaluated~\cite{Maier:2008he} (see also Ref.~\cite{Sturm:2008eb}).

In the perturbative calculation we renormalize the charm quark mass in the
$\overline{\rm MS}$ scheme. This enables us to extract directly the
corresponding short-distance quantity avoiding the detour to the pole mass and
the corresponding intrinsic uncertainty.

The results obtained for the charm quark mass
from equating the experimental and theoretical moments are
collected in Tab.~\ref{tab::mc1}.
In order to obtain these numbers we
set the renormalization scale to $\mu=3$~GeV and extract as a
consequence $m_c(3~\mbox{GeV})$. The uncertainties are due to
the experimental moments, $\delta\alpha_s(M_Z)=\pm0.002$, the variation
of $\mu$ between 2~GeV and 4~GeV and the non-perturbative gluon
condensate. 

In contrast to the corresponding table in Ref.~\cite{Kuhn:2007vp}
we included in Tab.~\ref{tab::mc1} the new four-loop results from
Ref.~\cite{Maier:2008he} for 
$n=2$. This leads to a shift in the central value from 0.979~GeV to
0.976~GeV. Furthermore the uncertainty of 6~MeV which was due to the 
absence of the four-loop result is removed.

\begin{table}[t]
\begin{center}
\tbl{\label{tab::mc1}Results for $m_c(3~\mbox{GeV})$ in GeV.
  The errors are from experiment,
  $\alpha_s$, variation of $\mu$ and the gluon condensate.
  The error from the yet unknown four-loop term is kept separate.
}
{\begin{tabular}{l|l|llll|l|l}
\hline
&&&&&&&\\[-.8em]
$n$ & $m_c(3~\mbox{GeV})$ & 
exp & $\alpha_s$ & $\mu$ & np & 
total & $\delta\bar{C}_n^{(30)}$ 
\\
\hline
        1&  0.986&  0.009&  0.009&  0.002&  0.001&  0.013&---\\
        2&  0.976&  0.006&  0.014&  0.005&  0.000&  0.016&---\\
        3&  0.982&  0.005&  0.014&  0.007&  0.002&  0.016&  0.010\\
        4&  1.012&  0.003&  0.008&  0.030&  0.007&  0.032&  0.016\\
\hline
\end{tabular}}
\end{center}
\end{table}

The results in Tab.~\ref{tab::mc1} show an impressive consistency
when going from $n=1$ to $n=4$ although the relative weight form the 
various energy regions contributing to ${\cal M}^{\rm exp}$ is completely
different: whereas for $n=1$ the region for $\sqrt{s}\ge5$~GeV amounts to
about 50\% of the resonance contribution it is less than 4\% for $n=3$.
Also the decomposition of the uncertainty changes
substantially as can be seen in Tab.~\ref{tab::mc1}. Whereas for $n=1$ the
contribution from the $\mu$ variation is negligible it exceeds the
experimental uncertainty for $n=3$.

\begin{figure}[t]
  \begin{center}
    \begin{tabular}{c}
      \leavevmode
      \epsfxsize=0.95\textwidth
      \epsffile{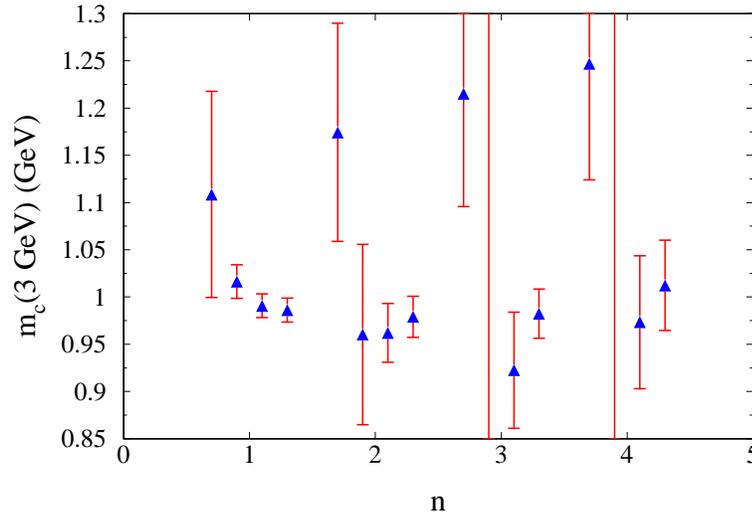}
    \end{tabular}
  \end{center}
  \vspace*{-2em}
  \caption{
    \label{fig::mom}$m_c(3~\mbox{GeV})$ for $n=1,2,3$ and $4$.
    For each value of $n$ the results from left to right correspond
    the inclusion of the one-, two-, three- and four-loop terms
    in the theory moments.
  }
\end{figure}

In Fig.~\ref{fig::mom} we show for the first four moments the result for
$m_c(3~\mbox{GeV})$ as a function of the loop order used for 
${\cal M}_n^{\rm th}$. One observes a nice convergence for each
$n$. Furthermore, the consistency among the three- and in particular the
four-loop results is clearly visible from this plot.

As final result of the analysis described in this Section we quote the value
given in Ref.~\cite{Kuhn:2007vp} which reads
\begin{eqnarray}
  m_c(3~\mbox{GeV}) &=& 0.986(13)~\mbox{GeV}
  \,.
  \label{eq::mc_1}
\end{eqnarray}


\section{\label{sec::ps}Lattice gauge theory and perturbative QCD}

In the recent years there has been a tremendous progress in developing precise
QCD simulations on the lattice. In particular, it has been possible to
simulate relativistic charm quarks using the so-called Highly Improved
Staggered Quark (HISQ) discretization of the quark
action~\cite{Follana:2006rc,Follana:2007uv}. In Ref.~\cite{Allison:2008xk}
this has been used to evaluate moments of the pseudo-scalar correlator 
with an uncertainty below 1\%.
The moments from the lattice calculation are equated with the ones
computed within perturbative QCD. In Ref.~\cite{Sturm:2008eb}
the second non-trivial moment could be evaluated with the help the
axial Ward identity from the  first moment of the longitudinal part of
the axial-vector current. Very recently this trick could be extended in order
to arrive at the third moment for the pseudo-scalar
current~\cite{Maier-ps-n3}.

Tab.~\ref{tab::mc_lat} summarizes the
results obtained for $m_c(3~\mbox{GeV})$ 
(for $n=2$ and $3$)\footnote{Note that for $n=1$ no charm quark mass
  can be determined since, in contrast to the vector correlator, the
  corresponding moment is dimensionless.}
together with the
corresponding uncertainties from the lattice, $\alpha_s$, 
missing higher order perturbative corrections and the gluon 
condensate.\footnote{For the presentation in this Section the notation of
  Ref.~\cite{Allison:2008xk} for
  the numeration of the moments has been translated to
  the one of Ref.~\cite{Kuhn:2007vp}.}

\begin{table}[t]
\begin{center}
\tbl{\label{tab::mc_lat}Results for $m_c(3~\mbox{GeV})$ in GeV.
  Both the total uncertainties are shown and the splitting into
  contributions from the lattice simulation,
  $\alpha_s$, missing higher order corrections and the non-perturbative gluon
  condensate.
}
{\begin{tabular}{l|l|llll|l}
  $n$ & $m_c(3~\mbox{GeV})$ & lattice & $\alpha_s$ & h.o. & np & total \\
  \hline
  2 & 0.986 & 0.008 & 0.003 & 0.004 & 0.003 & 0.010\\
  3 & 0.986 & 0.009 & 0.004 & 0.003 & 0.000 & 0.011 \\
\end{tabular}}
\end{center}
\end{table}

Like in the previous section we find also here an excellent agreement in the
central values which leads us to the final result
\begin{eqnarray}
  m_c(3~\mbox{GeV}) &=& 0.986(10)~\mbox{GeV}
  \,.
  \label{eq::mc_2}
\end{eqnarray}

Let us mention that
the dimensionless first moment can be used to extract a value for the
strong coupling. We can furthermore consider ratios of moments in order
to get rid of the overall dependence on $m_c$ and again extract $\alpha_s$. 
In Ref.~\cite{Allison:2008xk} this has been done for the ratio of the second
to the third moment which is known to four-loop order within perturbative QCD.
The two determinations lead to 
\begin{eqnarray}
  \alpha_s^{(4)}(3~\mbox{GeV}) &=& 0.251(6)
  \,,
\end{eqnarray}
which corresponds to\footnote{The calculation of the running and decoupling is
  easily done with the help of {\tt RunDec}~\cite{Chetyrkin:2000yt}.}
\begin{eqnarray}
  \alpha_s^{(5)}(M_Z) &=& 0.1174(12)
  \,.
\end{eqnarray}
This value agree well with the particle data group result~\cite{Yao:2006px} and other recent
determinations (see, e.g., Refs.~\cite{Baikov:2008jh,Kuhn-caqcd}).


\section{Summary}

In this contribution we have presented the two to date most precise
determinations of the charm quark mass. Let us stress once
again that, although in both cases moments of current correlators are
considered, the two methods rely on completely different experimental
input and on different theory calculations. Whereas in one case
perturbative QCD is compared with experimental data for $R(s)$, in the
second case high precision lattice simulations with dynamical charm
quarks are crucial ingredients.
It is quite impressive that the final results as given in
Eqs.~(\ref{eq::mc_1}) and~(\ref{eq::mc_2}) coincide both in the central value
and the uncertainty.

\begin{figure}[t]
  \begin{center}
    \begin{tabular}{c}
      \leavevmode
      \epsfxsize=\textwidth
      \epsffile{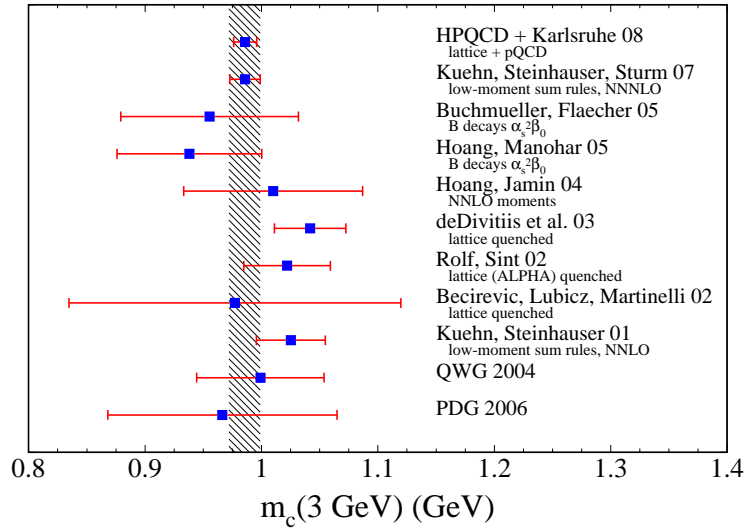}
    \end{tabular}
  \end{center}
  \caption{
    \label{fig::mc_compare}
    Comparison of recent determinations of $m_c(3~\mbox{GeV})$.
  }
\end{figure}

In Fig.~\ref{fig::mc_compare} we compare the results of
Section~\ref{sec::ve} and Section~\ref{sec::ps} with various other
recent determinations. One observes a good agreement, however, our
results are by far the most precise ones, as can be seen by the grey
band. 

Up to this point we have presented results for the $\overline{\rm MS}$ charm
quark mass evaluated at the scale $\mu=3$~GeV. In general, the comparison of
results from various analyses are performed for the scale-invariant mass,
$m_c(m_c)$ (see, e.g., Ref.~\cite{Yao:2006px}). Note, however, that the scale
$\mu=m_c$ is quite low and the numerical value of $\alpha_s$ is relatively
big. Thus, it would be more appropriate to perform the comparison at a higher
scale like $\mu=3$~GeV. Let us nevertheless present the scale-invariant charm
quark mass. From $0.986(10)$~GeV one obtains
\begin{eqnarray}
  m_c(m_c) &=& 1.268(9)~\mbox{GeV}\,.
\end{eqnarray}

The method described in Section~\ref{sec::ve} can also be used to extract the
bottom quark mass. The analysis of Ref.~\cite{Kuhn:2007vp} leads to
\begin{eqnarray}
  m_b(m_b) &=& 4.164(25)~\mbox{GeV}\,.
  \label{eq::mb}
\end{eqnarray}
After including the new four-loop results from Ref.~\cite{Maier:2008he} the
result of Eq.~(\ref{eq::mb}) becomes
\begin{eqnarray}
  m_b(m_b) &=& 4.162(19)~\mbox{GeV}\,,
  \label{eq::mb2}
\end{eqnarray}
which has a significantly reduced uncertainty.


\section*{Acknowledgments}
I would like to thank Konstantin Chetyrkin, Hans K\"uhn, Peter Lepage,
Christian Sturm and the HPQCD lattice group for a fruitful and pleasant
collaboration.
This work was supported by the DFG through SFB/TR~9.


\end{document}